\documentclass[11pt,a4paper]{article}
\usepackage[T1]{fontenc}

\usepackage[a4paper,top=2cm,bottom=2cm,left=3cm,right=3cm,marginparwidth=1.75cm]{geometry}
 
\usepackage{amsthm}
\usepackage{amsmath}
\usepackage{amssymb}
\usepackage{graphicx}
\usepackage{hyperref}
\usepackage{comment}

\newtheorem{theorem}{Theorem}
\newtheorem{proposition}[theorem]{Proposition}

\newtheorem{example}[theorem]{Example}

\usepackage[affil-it]{authblk}

\makeatletter
\renewcommand{\section}{\@startsection {section}{1}{\z@}%
              {24pt}{12pt} {\large\scshape\bfseries}}

\renewcommand{\subsection}{\@startsection {subsection}{2}{\z@}%
             {12pt}{12pt}  {\itshape\bfseries}}

\usepackage{apacite}
\usepackage{natbib}

\title{\bfseries \normalsize Randomized routing strategies of fleets of CAVs may prove market efficient}

\author[1]{Grzegorz Jamr\'oz*}
\author[1]{\L ukasz Gorczyca}
\author[1]{Rafa{\l} Kucharski}

\affil[1]{Faculty of Mathematics and Computer Science,  Jagiellonian University, Krak\'ow, Poland}
\affil[*]{Corresponding author, e-mail: grzegorz.jamroz@uj.edu.pl}

\date{\vspace{-5ex}}

\begin{document}
\maketitle

\section*{Short summary}\small
In future cities every driver may own a vehicle which could be either independently driven (HDV), or autonomously routed and piloted (CAV). The autonomous operations could be handled by a few competing companies. What is the market structure which would make this market aligned with city goals? In this paper we discuss a variant of the emerging market of collectively routed fleets of CAVs, where revenue for fleet operators is proportional to market share. We provide benchmark scenarios to compare the routing algorithms. We present several routing algorithms and demonstrate that, when the attitudes of human drivers towards CAVs exhibit significant diversity, randomised CAV routing, resulting in unpredictable travel times for HDVs, is more efficient than routing proportional to system optimum/user equilibrium. Based on this, we propose to improve the design of the market by augmenting the market-share objective with mean systemwide travel time in order to limit antisocial randomised strategies of fleet operators and drive the competition towards social welfare oriented cooperation.  

\textbf{Keywords}: Autonomous driving and routing, choice modelling, market competition, randomized routing

\section{Introduction}
Design of a fair, efficient and competitive market of urban CAV operations is likely to be a challenge, like for other markets,  compare \cite{kominers2017invitation}. The unfair predatory pricing policies of companies backed by huge venture capital like Amazon or Uber, e.g. \cite{khan2016amazon}, resulting in long-term consumer harming mono/duopolies, even though common, are undesirable and preventable. As the market share of CAVs in major cities such as Los Angeles and Beijing is still too small, e.g. \cite{GoldmanSachs2024AVForecast}, to significantly influence the urban routing market, we are in a position to design the market structure which would be most beneficial for cities and avoid learning from failures, see e.g. \cite{roth2018marketplaces}. 
Accordingly, in this paper we consider one variant of CAV autonomous-routing market where the revenue is derived only from market-share and not subscription/purchase cost. This potential future market is based on the following assumptions:
\begin{itemize}
\item Every driver owns a vehicle which can be either driven independently (HDV) or routed and navigated by one of several providers (CAV fleet operators). 
\item Autonomous driving is indistinguishable from human driving in traffic.
\item Each driver can freely choose/switch between HDV and one of CAV fleets. 
\item Fleet operators are remunerated collectively from traffic taxes proportionally to the number of users (market-share maximization).
\end{itemize} 
\begin{figure}[h!]
\centering
\includegraphics[scale=0.5]{"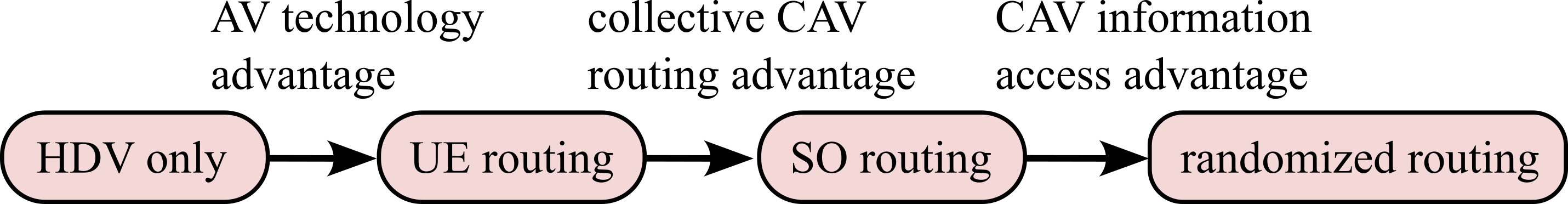"}
\caption{Incentives to switch to CAV. Compared to independent driving, AV technology offers better value of time for most drivers so that UE routing is enough to convince to switch to CAV. If this is not enough, leveraging collective market power via SO routing may convince some drivers. Finally, randomized routing which strips HDV drivers of the information which route will be fastest on a given day, may convince the most reluctant. }
\label{Fig_chrono}
\end{figure}
The above 'fee-free' design favours algorithmic routing and quality advancements over pure discount-based monopole building strategies which benefit from economy-of-scale effects, e.g. \cite{cramer2016disruptive}, and weak regulation, \cite{collier2018disrupting}. We study the dynamics of this market in a simplified case of one OD pair with two parallel routes. Fleet operators aim at maximum market share. Why would, however, a driver switch from independent driving (HDV) to autonomous driving and routing (CAV)? Three levels of incentives are presented in Fig. \ref{Fig_chrono}:
\begin{enumerate}
\item[i)] People's travel time cost spent in an AV is less than in an HDV. This is likely to be the case for most drivers, see \cite{de2019impact}. These customers are 'taken for free' by fleet operators by using user equilibrium (UE) routing.
\item[ii)] Collective routing can improve average travel times compared to user equilibrium by routing according to System Optimum (SO) with turn-taking for equity, see \cite{hoffmann2025wardropian} and references therein. 
\item[iii)] SO routing is unstable as every driver may be inclined to defect and use the faster route. Moreover, drivers originally reluctant to use AVs will still not be convinced. Randomized routing addresses these two issues by deliberately introducing uncertainty and depriving independent drivers of the information which route is likely to be fastest on a given day. 
\end{enumerate}

To study the above mentioned mechanisms and design a reasonable market structure, we:
\begin{itemize}
\item Introduce a framework(benchmark) for comparing algorithms for routing fleets of CAVs aimed at maximizing market share, including both baseline algorithms and more sophisticated heuristics.
\item Introduce dynamic randomized algorithms  for routing and demonstrate experimentally that they are strictly superior to non-randomized algorithms; note that randomization in the static full market-share context was discussed in \cite{jamroz2025market}.
\item Show that including average system travel time minimization in the objective renders randomized algorithms less efficient and favors algorithms more aligned with city goals, while preserving competition.
\end{itemize}

\section{Methodology}

\subsection{System dynamics}
\begin{figure}
\centering
\includegraphics[scale=0.45]{"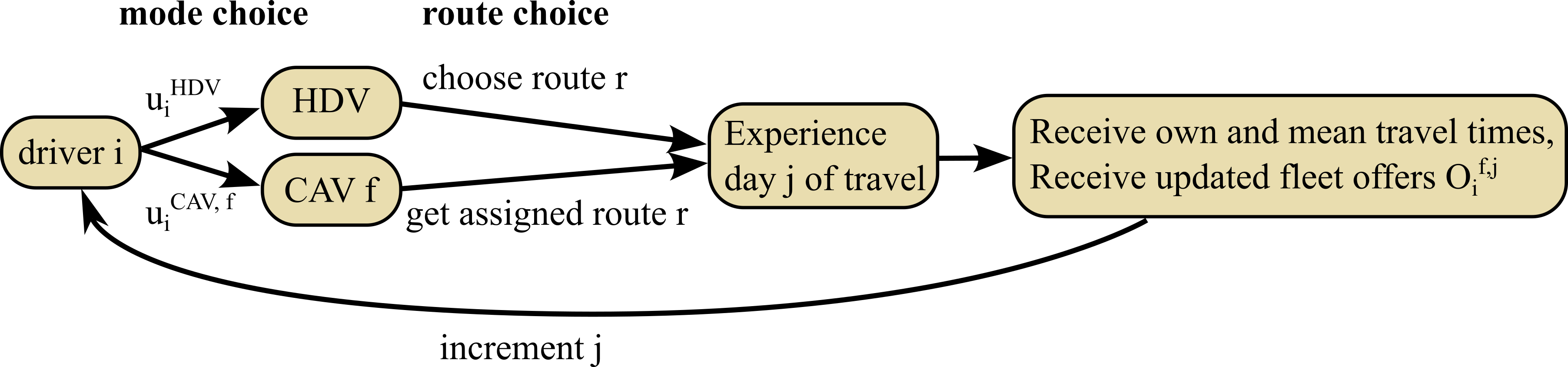"}
\caption{Dynamics of the system at a glance. Every day each driver $i$ selects the mode (HDV or CAV f) and route based on utilities derived from experienced and mean travel times and fleet operators' travel time offers.}   
\label{Fig_overview}
\end{figure}

We present benchmark scenarios modelling the competition of fleet operators in systems consisting, for simplicity, of one OD pair and parallel routes $r = 1,\dots, R$. We consider a fixed number $N_D$ drivers, who every day decide whether to drive independently or join one of $N_{FLEET}$ fleets of autonomously routed vehicles, based on travel times offered by each of the fleets and credibility of fleet offers. Consequently, the system can be considered a game with $N_D + N_{FLEET}$ players, who make decisions as follows. On a given day $j$, sequentially, see Fig. \ref{Fig_overview}:
\begin{enumerate}
\item [i)] Each fleet operator $f \in \{0,1 \dots,N_{FLEET}-1\}$ creates an offer vector $O^{f, j}_i$ for $i = 0,1,\dots, N_D-1$ based on some algorithm $\bf{Algo^f_{offer}}$ such that $O_i^{f,j}$ is the mean long-term travel time fleet $f$ offers to driver $i$.  
\item [ii)] Each driver $i = 0,1,\dots, N_D-1$ decides, based on previous travel experience and offers from fleet operators whether to remain independent (HDV) or be part of one of the fleets, resulting in mode vector $M^j \in \{HDV, 0, 1, \dots, N_{FLEET}-1\}^{N_D}$, where $M^j_i$ is the the mode chosen by driver $i$ on day $j$. 
\item [iii)] Once the drivers have committed to fleets or to staying independent HDVs, each fleet $f$ creates, based on some algorithm $\bf{Algo^f_{routing}}$, a routing vector $[r^{f, j}]_i$ for drivers $i \in I^{f,j} := \{i: M_i^j = f\}$, i.e drivers who have chosen fleet $f$ on day $j$. The value $r^{f, j}_i \in \{1,\dots, R\}$ corresponds then to the route via which driver $i$ will be routed on day $j$. The drivers $i$ who have decided to remain independent, choose routes based on minimizing disutility, see below, resulting in vector $r^{HDV,j}_i$. The resulting routing vector on day $j$ is given by $r_i^j = r_i^{M_i^j, j}$ such that $r_i^j$ is the route via which driver $i$ will go on day $j$. 
\end{enumerate}

After all the routing decisions have been made, a day of travelling is simulated, resulting in route travel time $t^j_r$, for $r = 1, \dots, R$. The travel times are assumed to be independent of driver identity, however they depend on congestion and in our considerations are given by the BPR function:

\begin{equation}
t^{BPR} (q) = (5 min)*\left(1 + (q/(0.5*N_D))^2\right)\label{eq_BPRA}
\end{equation}
where $q$ is the number of vehicles driving via the route on a given day. 

\subsection{Human behaviour}
We assume that each driver $i$ has, for each fleet $f$, a discount factor $\gamma_i^{F,f} > 0$, which models the attitude towards using fleet $f$. This attitude can be decomposed into a part corresponding to the general liking of using fleets, $\gamma_i^F$ (drawn from a given known distribution, in our simulations assumed to be gaussian, clipped at $0$, centred at $0.7$ with standard deviation $0.2$, compare \cite{de2019impact}) and a factor specific to a given fleet $\gamma_i^{F,specf}$, drawn from $N(0,(0.15)^2)$ (gaussian centred at $0$ and standard deviation $0.15$), resulting in $\gamma_i^{F,f} = \gamma_i^F + \gamma_i^{F,specf}$. The utility of driving independently $u_i^{HDV}$ is given by the mean travel times over the past $j_{avg}$ days, sampled uniformly from $\{1,2,\dots,9\}$, weighted with logit probabilities. Contrariwise, the utility $u_i^{CAV,f}$ of using fleet $f$ is proportional to the travel time offered multiplied by the discount factor $\gamma_i^f$ and divided by credibility . Consequently, on day $j$, for $\overline{t_r} = \sum_{\iota = j-j_{avg}}^{j-1} t^{\iota}_r$, disutilities of using different modes are:
\begin{eqnarray}
u_i^{HDV} &=& \sum_{r \in R} e^{-\beta \overline{t_r}} \overline{t_r}/ \sum_{r \in R} e^{-\beta \overline{t_r}} \label{eq_utHDV}\\
u_i^{CAV,f} &=& \gamma^{F,f}_i * O_i^{f,j} / Cred_i^{f} \label{eq_utCAV}, 
\end{eqnarray}
where $O_i^{f,j}$ is the travel time offered by fleet $f$ to driver $i$ and $Cred^{f}_i >0$ is the credibility of fleet $f$ to driver $i$. The initial credibility of fleets is assumed to be $1$ and it is updated by 
\begin{equation}
\label{eq_cred}
Cred_i^f \leftarrow (1 - \alpha)*Cred_i^f + \alpha * O_i^{f,j} / t_{r_i^{j}}^{j}
\end{equation}
for some update rate $\alpha$ (which in this paper is assumed to equal $0.2$), if driver $i$ chose fleet $f$ on the previous day $j$ and was routed via $r_i^j$. This means that human drivers are inclined to try out different fleets initially, provided the (honest or dishonest) travel time offer $O_i^{f,j}$ is attractive enough. However, if the offers are later not confirmed in the real driving, the credibility of a given fleet decreases. The gradual update of credibility is important as the fleets are supposed to offer long-term average travel times and not exact travel times on the following day. 

Finally, driver $i$ is assumed to choose the mode (HDV or one of the fleets $f$) with lowest disutility. If HDV is chosen, the route choice is based on logit choice of routes, i.e. probability of choosing route $r$ is given by $ e^{-\beta \overline{t_r}} / \sum_{r \in R} e^{-\beta \overline{t_r}}$, where we assume $\beta = 0.2$. The rationale for choosing this model is that in the future even HDVs are likely to use some apps to propose routes, and a simple way to choose roughly optimal routes on a population level with small error is to use logit probabilities with small $\beta$, while mode choice, which would perhaps follow some bounded rationality model with near-deterministic logit probabilities can be approximated by deterministic mode choice.

\subsection{Algorithms for CAV fleet behaviour}
\subsection{Algorithm 1: Proportional SO router}
This baseline algorithm offers routing aimed at achieving System Optimum. To reach this goal it first determines the system optimal flows $\bold{q^{SO}} = (q^{SO}_1, \dots, q^{SO}_r)$, which is straightforward in the case of two parallel routes as well as the average travel time at System Optimum, $t^{SO}_{avg}$. 
$\bf{Algo^f_{offer}}$ is based on offering every driver average system optimal travel time, i.e.
\begin{equation*}
O_i^{f,j} = t^{SO}_{avg}
\end{equation*}
for every day $j$ and driver $i$. $\bf{Algo^f_{routing}}$ is the following. Given the set of drivers $I^{f,j}$ 
 that have chosen fleet $f$ on day $j$ the drivers are split randomly with probabilities proportional to $\bold{q^{SO}}$. 
In the case of two available routes, it consists of selecting a uniformly random sample $I^{rand}$ of $\left\lfloor |I^{f,j}| * q^{SO}_1/|\bold{q^{SO}}| \right \rfloor$ agent indices from $I^{f,j}$ and setting 
\begin{equation*}
r^{f,j}_i = 
\begin{cases}
0 &\mbox{ for } i \in I^{rand},  \\
1 &\mbox{ for } i \in I^{f,j} \backslash I^{rand}.
\end{cases}
\end{equation*}

The rationale behind this algorithm is that if all the drivers choose this fleet, routing according to system optimal assignment is the most efficient way to minimize average travel time, which is desirable. The downside is that this routing treats all the drivers in the same way and disregards different attitudes towards the fleet, making the routing potentially inefficient. Furthermore, in the case of non-full market penetration, the offered travel times cannot in fact be realized, which decreseas the credibility of the fleet.  A variant of the algorithm consists in offering travel times lower than system optimum, i.e. 
\begin{equation}
\label{eq_scaled_SO_offer}
O_i^{f,j} = \kappa t^{SO}_{avg}
\end{equation}
for some $\kappa \in (0,1]$ in order to win over customers at the beginning. 

\subsection{Algorithm 2: Proportional UE router}
This algorithm is analogous to Proportional SO routes, with $\bold{q^{SO}}$ replaced by $\bold{q^{UE}}$, i.e. user-equilibrium flows and setting $O_i^{f,j} = t^{UE}$, where $t^{UE}$ is the travel time via both routes at User Equilibrium. We note that in the case of two equivalent routes, SO and UE coincide and there is no distinction between the two algorithms.

\subsection{Algorithm 3: Randomized algorithm RFlexV}
The standard offer algorithm $\bf{Algo^f_{offer}}$ is to offer the scaled average system optimal travel time \eqref{eq_scaled_SO_offer} with $\kappa = 0.5$ (by default). The routing is given by the following randomized $\bf{Algo^f_{routing}}$ algorithm. Let $|I^{f,j}|$ be the number of drivers who have chosen fleet $f$ on day $j$. For two equivalent routes $0,1$ the fleet operator proposes a symmetric randomized strategy such that 
\begin{itemize}
\item all the current members of the fleet are happy (i.e. will not want to leave the fleet to become an HDV on the next day),
\item the strategy is as randomized as possible in order to make independent routing difficult for other drivers and coax them to switch to the fleet. 
\end{itemize}
To achieve this, we assume that the remaining drivers will split 50-50 (as the routing is symmetric) and choose
\begin{equation}
\label{eq_nfasterstar}
n^{faster*} \in \arg\min\left\{ N_{unhappy}(n_{faster}): n_{faster} = 1,2,\dots, \lfloor|I^{f,j}|/2\rfloor\right\},
\end{equation}
where $N_{unhappy}(n_{faster})$ stands for the number of drivers who would rather switch to HDV than stay with the fleet when the fleet puts $n_{faster}$ agents with highest discount factors on the faster route and is given by
\begin{equation*}
N_{unhappy}(n_{faster}) = \left|\left\{i \in I^{f,j}: \gamma_i^{F,f} t^{sim}_{r^{f,j}_i}  > \overline{t_r^{sim}}\right\}\right|,
\end{equation*}
where $\overline{t_r^{sim}}:=\frac 1 {N_{R}} \sum_{r=1}^{N_{R}} t^{sim}_r$ and $t^{sim}_r$ is the simulated travel time of route $r$ provided $n_{faster} < |I^{f,j}|/2$ fleet members are routed, without loss of generality, via route $0$ (which will be the faster route) and  $|I^{f,j}| - n_{faster}$ drivers are put on the other (slower) route $1$. We obtain:
\begin{eqnarray*}
t^{sim}_0 &=& t_0\left(n_{faster} + \lfloor(N_D - |I^{f,j}|)/2\rfloor\right)\\
t^{sim}_1 &=& t_1\left(|I^{f,j}| - n_{faster} + \lceil(N_D - |I^{f,j}|)/2\rceil\right)
\end{eqnarray*}
and $t_0 = t_1$ are the known functional dependences of travel time on flow, see \eqref{eq_BPRA}. 

Now, let $n^{faster*}$ be the smallest such that equation \eqref{eq_nfasterstar} is satisfied. The algorithm assigns $n^{faster*}$ drivers with highest discount factors to a random route and $|I^{f,j}| - n^{faster*}$ drivers to the alternative route, i.e. for $r^*$ - a random number from $\{0, 1\}$ we have
\begin{equation*}
r_i^{f,j} = 
\begin{cases}
r^* &\mbox{ if } s(i) < n^{faster*},\\
1-r^* &\mbox{ otherwise, }
\end{cases}
\end{equation*}
where 
\begin{equation}
\label{eq_sorting_fun}
s: \{0,1,\dots, |I^{f,j}|-1\} \to I^{f,j}
\end{equation}
is a sorting function such that $\gamma^{F,f}_{s(0)}, \dots, \gamma^{F,f}_{s(|I^{f,j}|)}$ is a non-decreasing sequence (note $s$ depends on $I^{f,j}$). 
The algorithm tries to maintain its current customer base while randomizing the assignment so that it is less convenient to remain an independent driver for discount factors $\gamma^{F,f}_i > 1$. 
This, however, (even in the case of one fleet) may be suboptimal, as a driver $i$ with $\gamma_i^{F,f}>1$ does not have to be routed via the faster route every day, but for instance only on $80\%$ of days, see \cite{jamroz2025market}. 

\subsection{Algorithm 4: RFlex}
RFlex is an extension of RFlexV, optimizing the choice of drivers to be routed via the faster route so that drivers who do not need to be routed via the faster route are not routed via it. To this end, it computes the least necessary share $sh_i$ of routing driver $i$ via the faster route to keep $i$ happy as:
\begin{equation*}
sh^*_i = \min\left\{s: \gamma^{F,f}_i (s t_0^{sim} + (1-s)t_1^{sim}) <  \overline{t_r^{sim}} \right\}. 
\end{equation*}
Solving the above equation we obtain:
\begin{eqnarray*}
sh_i^* = \frac{t_1^{sim} - \overline{t_r^{sim}} / \gamma_i^{F,f}}{(t_1^{sim} - t_0^{sim})} 
\end{eqnarray*}
with the caveat that if $sh_i^* < 0$ then driver $i$ will always be happy (we put $sh_i^* = 0$) and when $sh_i^* > 1$ then driver $i$ cannot be made happy (we put $sh_i^* = \infty$, because this driver cannot be convinced anyway for the given $n_{faster}$). Let $s(i)$ be the sorting function from \eqref{eq_sorting_fun}. 
The theoretical maximum number of drivers that can be made happy is given by 
\begin{equation*}
n_{maxhappy} = max\left\{n: \frac 1 n\sum_{i=0}^{n-1} sh^*_{s(i)} < \frac {n_{faster}}{n}\right\}.
\end{equation*}
The routing should theoretically now be given by randomly assigning the drivers to routes by $r^{f,j}_i$ such that  
\begin{equation}
\label{eq_expsh}
\mathbb{E} r^{f,j}_{s(i)} < sh^*_{s(i)}
\end{equation}
for every $i = 0, \dots, n-1$, where, without loss of generality $0$ is the faster route and $1$ is the slower route.  The overall routing could consist in finding $n^{faster}$ such that $n_{maxhappy}$ is the highest and routing such that \eqref{eq_expsh} is satisfied. This would work if human drivers fully trusted the fleet operator to deliver the offered travel times. However, in practice, the drivers might be discouraged if they are not offered good travel times initially after joining the fleet and, moreover, the number of participants of the fleet changes making the theoretical delivery of offered travel times tricky. In view of the above, we opt for a heuristic $\bf{Algo^f_{routing}}$ given by setting the target ratio of routing every driver via the faster route as $sh_i^{target} = sh_i^* / \sigma$, where $\sigma \in (0,1]$ is a parameter to optimize (we use $\sigma = 0.4$). We obtain:
\begin{equation*}
r_i^{f,j} = \begin{cases}
r^* &\mbox{ if } \frac {d_i^{faster}}{d_i^{faster} + d_i^{slower} + 1} > sh_i^{target},
\\ 
1 - r^* &\mbox{ otherwise,} 
\end{cases}
\end{equation*}
where $d_i^{faster}$ and $d_i^{slower}$ is the number of days driver $i$ was routed via the faster (slower) alternative since joining the fleet and $r^*$ is a random number from $\{0,1\}$. 

\subsection{Estimation of human drivers' discount factors}
Across most scenarios, we assume that the discount factors are known to the fleets. 

\section{Results and discussion}
In this paper we consider the simplest benchmark scenario of two equivalent routes and discount factors known to fleet operators, leaving extensions with more realistic details to further work.
The parameters used are listed in Table \ref{Tab_scenario_params}. 
\begin{table}[h!]
\centering
\caption{Simulation parameters}
\label{Tab_scenario_params}
\begin{tabular}{|c | c | c |}
\hline
Parameter & Symbol & Value\\
\hline
Number of drivers & $N_D$ & $200$\\
Number of days & $J$ & $300$\\
Route 0 delay function & $t_0(q_0)$ & $t^{BPR}$\\
Route 1 delay function & $t_1(q_1)$ & $t^{BPR}$\\
Initial fleets' credibility & $Cred_{init}$ & $1$\\
Fleet discount factor distribution & $\gamma^{FD}$ & $N(0.7, (0.2)^2) $\\
Fleet specific discount factor term & $\gamma^
{F,specfD}$ & $N(0, (0.15)^2)$\\
Discount factors known to Fleets? & $DiscF_{known}$ & True\\ Discount factor distribution known? & $DiscD_{known}$ & True\\
HDV logit parameter & $\beta$ & $0.2$\\
\hline
\end{tabular}
\end{table}

\begin{table}[b]
\caption{Algorithms used}
\label{Tab_Algos}
\begin{tabular}{ | c | c | c |}
\hline
Algorithm & Offer creation $\bf{Algo^f_{offer}}$ & Assignment method $\bf{Algo^f_{routing}}$\\
\hline
SO & Avg SO travel time & Proportional to SO + random turn-taking\\
SO- & $0.8 * $ Avg SO travel time & Proportional to SO + random turn-taking\\
UE & Avg UE travel time & Proportional to UE\\
UE- & $0.5 * $ Avg UE travel time & Proportional to UE\\
RFlexV & Avg SO travel time & highest $\gamma^F$ to (randomized) faster route\\
RFlexV- & $0.5 * $ Avg SO travel time & highest $\gamma^F$ to (randomized) faster route\\
RFlex & Avg SO travel time & Adaptive randomized routing to make happy\\
RFlex- & $0.5 * $ Avg SO travel time & Adaptive randomized routing to make happy\\
Infty & Infinite & No assignment, used for one-fleet scenarios\\
Your Algorithm & ? & ?\\
\hline
\end{tabular}
\end{table}

The execution of the benchmark scenarios proceeds as follows.
\begin{enumerate}
\item A list of $N_D$ human drivers is created. At the beginning, each driver $i \in \{0,\dots,N_D -1\}$ has the initial credibility towards the fleets equal $Cred_{init}$ and a couple of discount factors $\gamma^{F,0}_i, \gamma^{F,1}_i$, where $\gamma^{F,f}_i = \xi_i + \xi_{i,f}$ for $f = 0,1$ where $\xi_i$ a number drawn from $\gamma^{FD}$ and $\xi_{i,f}$ drawn independently from $\gamma^{F,specfD}$ for $f=1,2$. If any of the resulting factors is $\le 0$ then it is set to $\varepsilon$, where $\varepsilon = 0.0001$ is a very small positive number. In this way, the attitude of a human towards fleet $f$ is broken down into the general attitude towards using fleets of AVs and fleet specific components.  
\item Fleets $f$ for $f \in \{0,1\}$ are created. Every fleet uses one of the algorithms specified in Table \ref{Tab_Algos} and described in detail in Section {\it Methodology - Algorithms for CAV fleet behaviour}.
\item The dynamics of the system are simulated in line with Section {\it Methodology - System Dynamics}. Various parameters are recorded.
\item The parameters like market share of a fleet are plotted and compared to other algorithms. 
\end{enumerate}

\subsection{Experiment 1: Offering realistic travel times vs. unrealistically short travel times}
\begin{figure}[h]
\label{Fig_1}
\hspace*{-0.5cm}
\includegraphics[scale=0.6]{"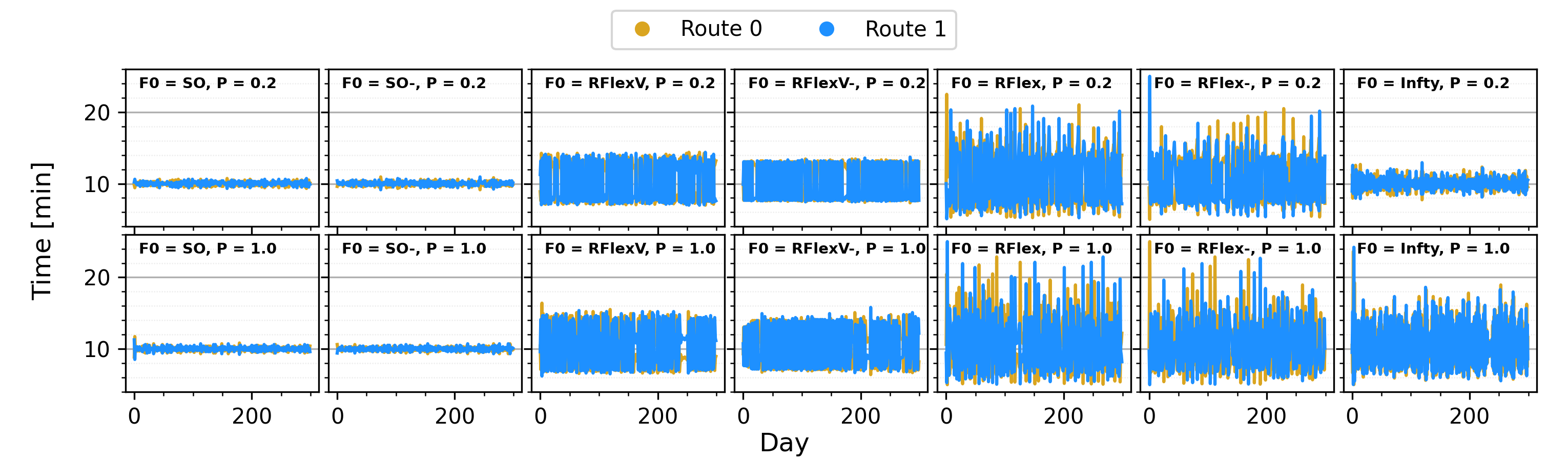"}
\hspace*{-0.5cm}
\includegraphics[scale=0.6]{"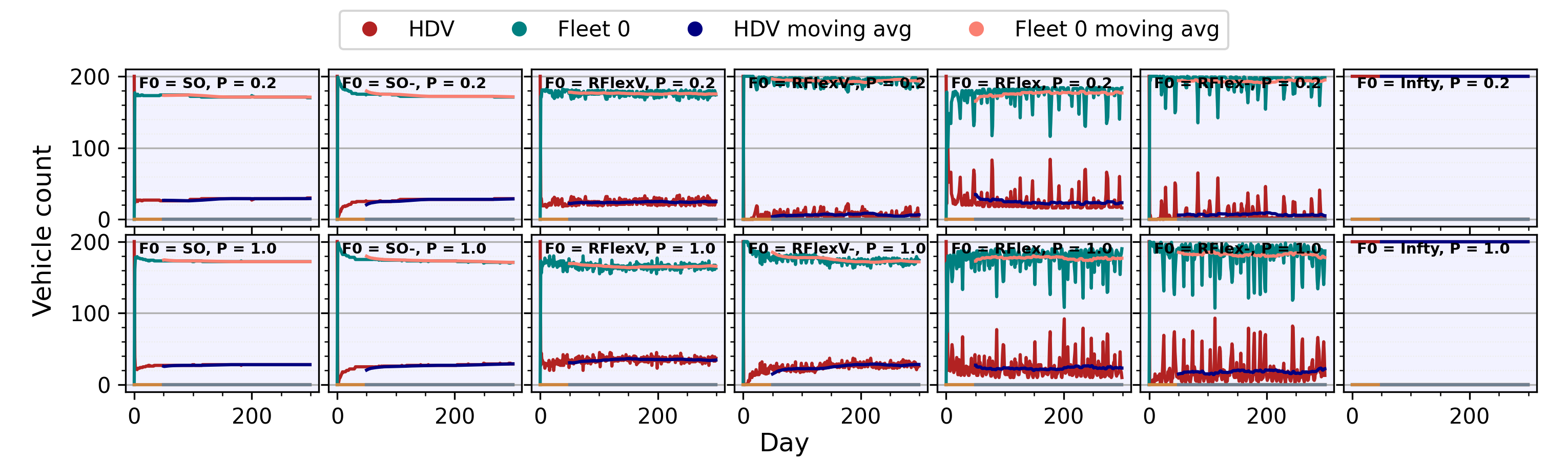"}
\caption{Comparison of properties of the system for different logit parameters of human driver assignment (P) and routing algorithms (F0) with travel time offers either unrelistically low (with '-') or more realistic (without '-'). Upper panel: Logit parameter $0.2$ is a preferred choice for the human population as it results in relatively stable assignment when Fleet is absent (F0 = infty). Lower panel: The performance of '-' algorithms is superior to those offering more realistic travel times. }
\end{figure}

Fig. \ref{Fig_1} shows the performance of different algorithms in the variants with best realistically possible mean travel times offered (without '-') and with offered travel times much faster than achievable (with '-'). For SO the advantage for '-' only persistis temporarily whereas for RFlex and RFlexV it is long-term. Moreover the logit parameter of human assignment equal $0.2$ offers better, stable travel times, which is preferred over every driver striving to choose the faster route every time which results in high variance (P = 1.0). Therefore, for further experiments we choose to set the logit parameter to $0.2$ and study only algorithms offering unrealistically short travel times. 

\subsection{Experiment 2: Randomized algorithms vs. baseline algorithms}
\begin{figure}
\hspace*{-0.5cm}
\includegraphics[scale=0.6]{"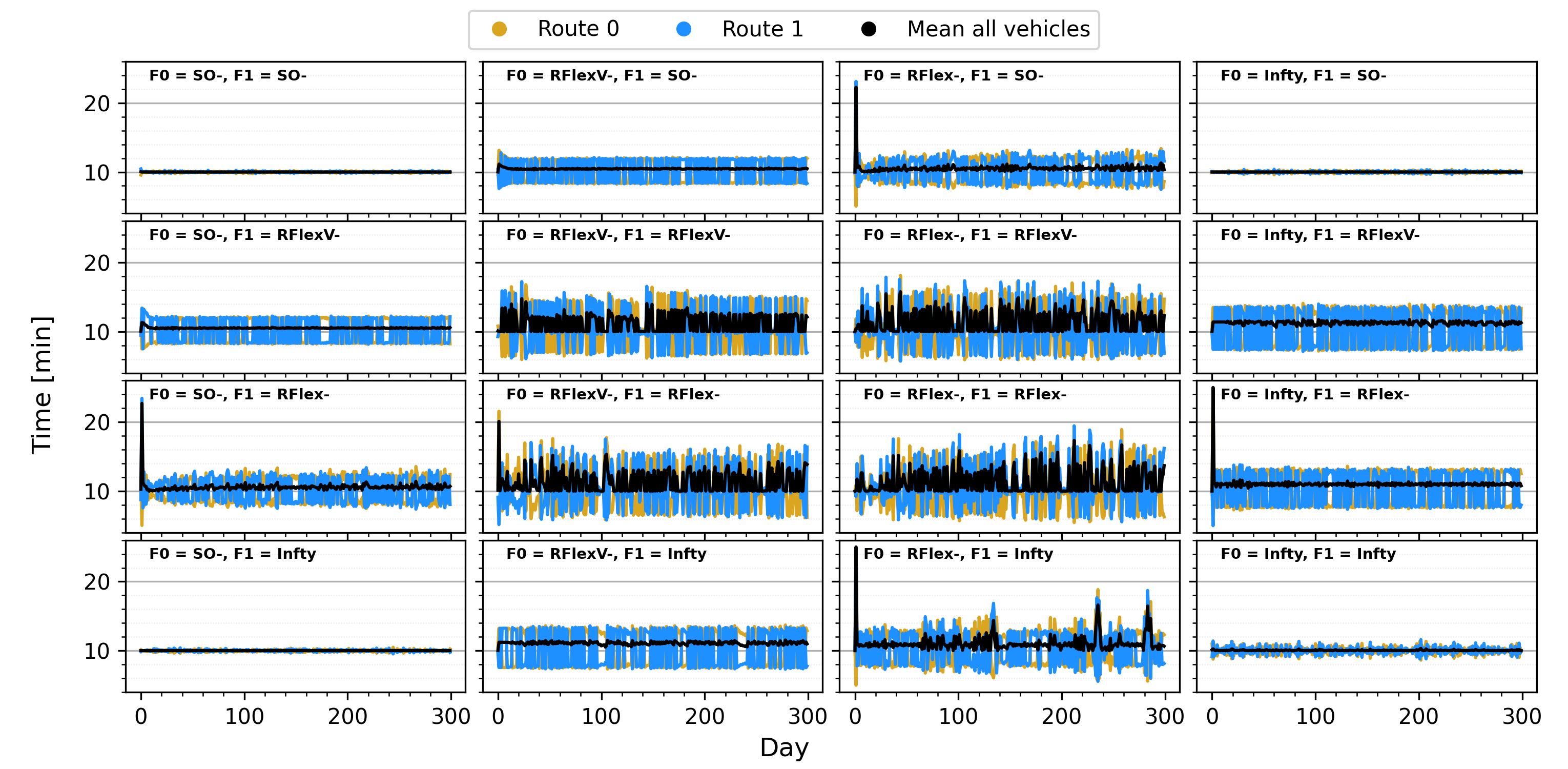"}
\hspace*{-0.5cm}
\includegraphics[scale=0.6]{"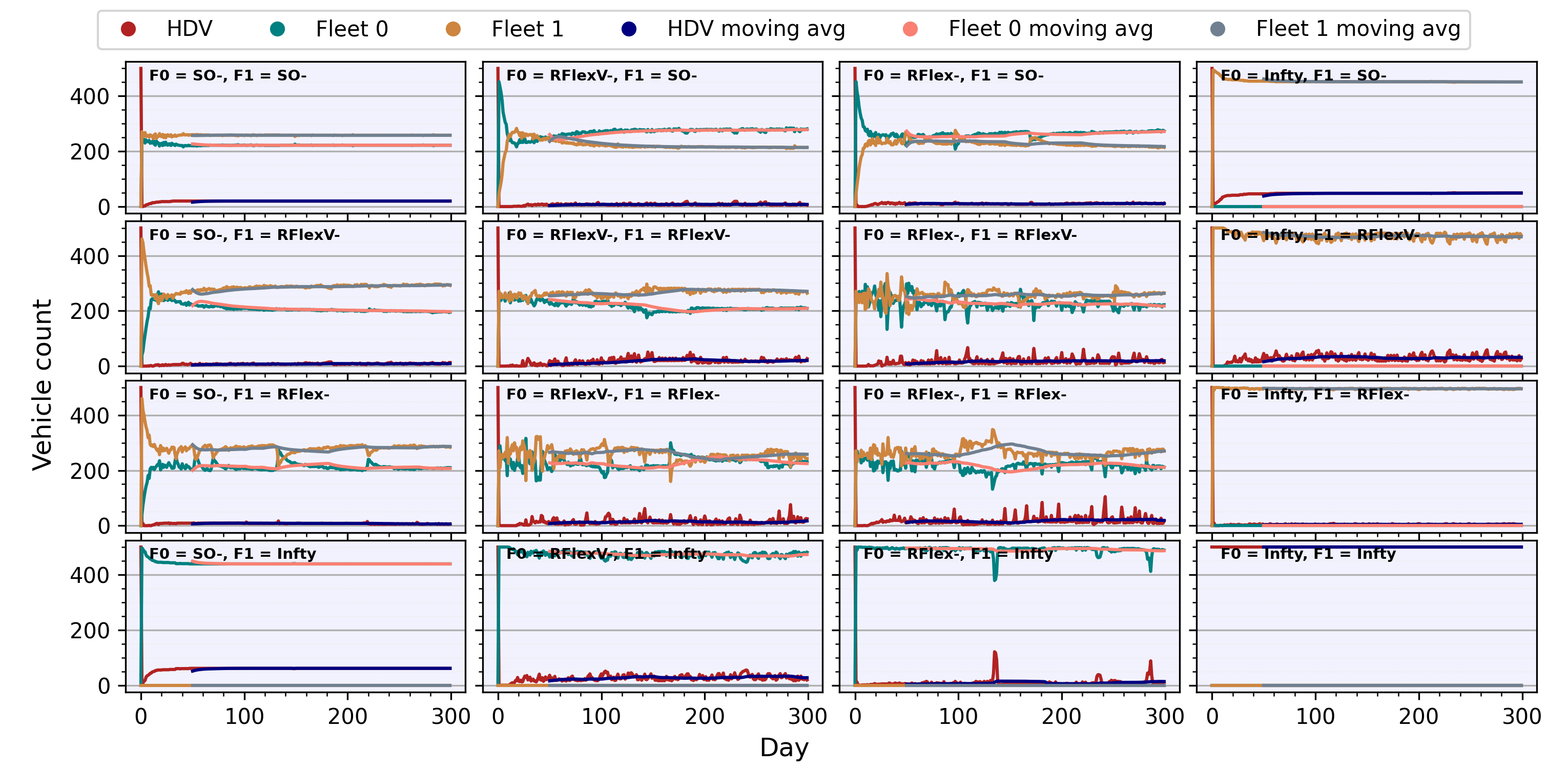"}
\caption{Competition of fleet operators in a benchmark scenario. Randomized algorithms RFlexV- and RFlex- perform better in one-fleet scenario (against F1 = Infty), while in two-fleet scenarios the results are mixed. The travel times when one or two fleets use randomized algorithms are highly variable. }
\label{Fig_3}
\end{figure}

\begin{figure}
\hspace*{-0.5cm}
\includegraphics[scale=0.6]{"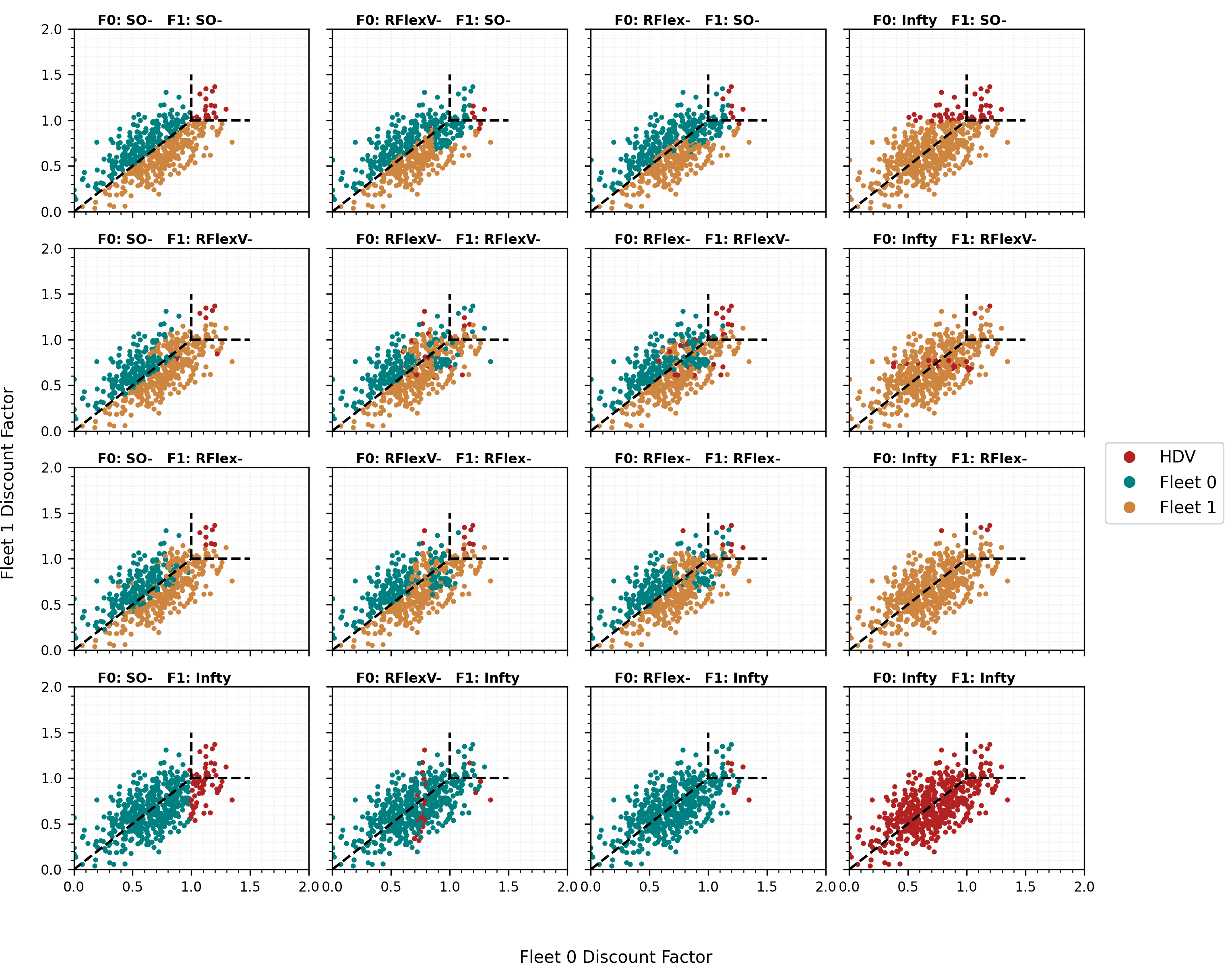"}
\caption{Usage of HDV or Fleet 0 or Fleet 1 on day $150$ in simulations. Every dot in a panel corresponds to one vehicle and is plotted at coordinates corresponding to respective discount factors of fleets. Dashed lines separate the natural basins of belonging - as in the upper left panel. More efficient algorithms are able to attract and retain customers from outside of their basins. The HDV sample of discount factors is the same in every panel.}
\label{Fig_3bis}
\end{figure}

In Fig. \ref{Fig_3} we pitch the algorithms against each other. The outcomes reveal that RFlexV is superior to SO and RFlex is superior to RFlexV when there is one fleet (last row/last column in bottom panel). However, when there are two fleet operators, the results are mixed, with RFlex/RFlexV performing slightly better than SO against SO (first row/first column, bottom panel), however not necessarily so against either RFlexV or RFlex (four central subplots). The upper panel demonstrates that randomized strategies entail large variations of travel times and the increase of mean travel time in the system. Fig. \ref{Fig_3bis} shows which vehicles belong to which fleet on a sample day 150 plotted against discount factors towards both fleets. The baseline natural split is for both fleets using SO algorithms - drivers join the fleet with better discount factor provided it is $<1$. This changes when more advanced algorithms are used. E.g. in the last row it can be seen how RFlexV/RFlex can convince even drivers who, without randomization would not join the fleet. Moreover, if a fleet uses a randomized algorithm then the other fleet sticking to the baseline SO algorithm fails to be competitive (first row and left column) and has to switch to a different algorithm. 
The four central panels illustrate  continual competition for customers. 
The sample of discount factors is the same for every subplot. 

\subsection{Experiment 3: SO component in objective}
Experiment 2 shows that randomized algorithms are efficient in reaching higher market share, however for the price of increased and highly varying average travel times. Here, we demonstrate that a combination of share maximization with minimization of average travel time could be an objective which would be more socially acceptable (less oscillations), while preserving competition and discouriging randomization.  
\begin{figure}[h!]
\includegraphics[scale=0.6]{"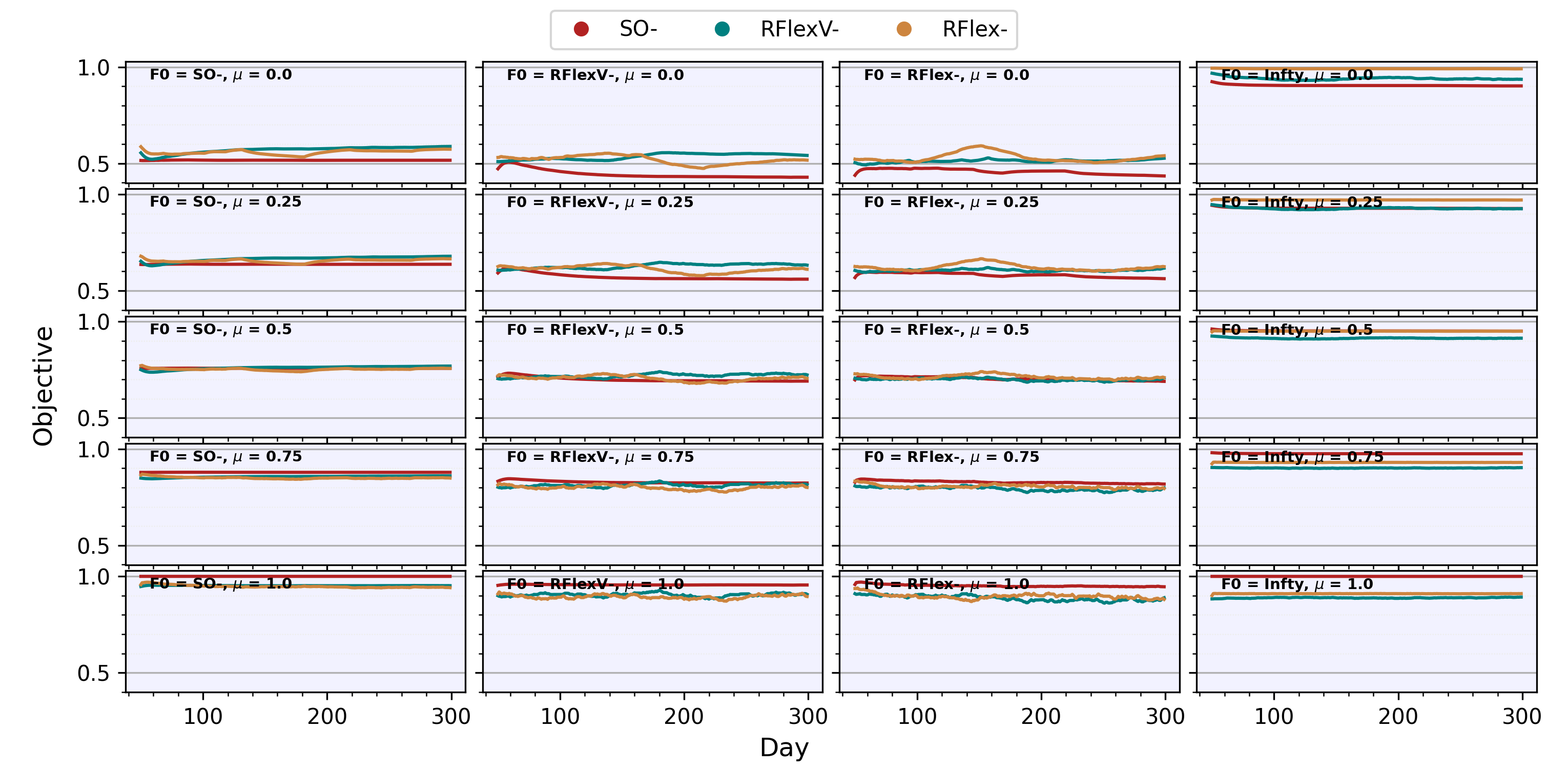"}
\caption{Smoothed-out objective (50-day moving average) for different values of $\mu$ and algorithm used by the fleet $F0$ against the fleet $F1$ is competing. Red/green/orange line correspond to $F1$ using SO-/RFlexV-/RFlex-, respectively. With growing $\mu$ the algorithm aiming at SO, ie. SO- becomes more and more competitive and eventually superior to randomized algorithms aiming solely at share maximization. }
\label{Fig5}
\end{figure}
To this end let
\begin{eqnarray*}
t^j_{avg} = \frac 1 {N_D} \sum_{r \in R} q_r^j t_r^j
\end{eqnarray*}
where $q_r^j$ is the flow (number) of vehicles on route $r$ and day $j$. Let $\tau^j_{avg} := t^{SO}_{avg}/t^j_{avg}$ be the normalized average travel time and let 
$n^{f,j} = |I^{f,j}|/N_D$ be the share of fleet $f$ on day $j$. Fig. \ref{Fig5} shows the objectives we propose (to be maximized):
\begin{equation*}
Obj^{\mu} = (1-\mu)n^{f,j} + \mu \tau^j_{avg} 
\end{equation*}
for different values of parameter $\mu$. We note that $\mu = 0$ corresponds to market share maximization while $\mu = 1 $ corresponds to minimizing average travel time. As already noticed before, for $\mu=0$ the randomized algorithms $RFlex-$ and $RFlexV-$ are superior to SO oriented $SO-$. This changes with increasing $\mu$ and for $\mu \approx 0.5$ the plain algorithm $SO-$ becomes competitive enough and randomization is discouraged; indeed using randomization resulting in increased average travel times reduces the objective, i.e. payoff for the fleet operator.

\section{Conclusions}
In the paper we introduced a benchmark for comparing routing algorithms competing for customers in the scenario with diverse drivers. We provided baseline algorithms (SO) as well as more advanced heuristic randomized algorithms (RFlexV, RFlex). Running the benchmark scenario we found that:
\begin{itemize}
\item Dynamic routing algorithms using randomized strategies, which result in highly variable travel times seem to be more efficient than baselines when the sole objective is maximization of market share (and attitudes of drivers towards fleets are known). 
\item If one fleet uses a randomized algorithm, the other cannot stick to the baseline SO routing to stay competitive. 
\item The city could shape the market by setting the remuneration scheme based on combining the market share objective and deviation from System Optimum objective.
\item There are indications that cities might be able to preserve competition while discouraging unwanted market-share optimizing behaviours such as randomized routing. 
\end{itemize}

Future work will encompass: 
\begin{itemize}
\item Including benchmark scenarios with more routes than $2$ and more fleets than $2$.
\item Developing even more efficient routing algorithms which directly address competition against other fleet providers as well as take into account the objective set by city authorities.
\item Developing ML-based routing algorithms in the competition case.
\item Considering scenarios where fleet discount factors are not known to fleet operators and have to be inferred.
\item Simulations in realistic city-scale simulators such as open-source SUMO, \cite{SUMO}. 
\end{itemize}

\section*{Acknowledgements}
This work was financed by the European Union within the Horizon Europe Framework Programme (ERC Starting Grant COeXISTENCE no. 101075838). Views and opinions expressed are however those of the authors only and do not necessarily reflect those of the European Union or the European Research Council Executive Agency. Neither the European Union nor the granting authority can be held responsible for them.

\bibliography{references}

\appendix

\section{Appendix: SUMO-based simulations}
\begin{figure}[h!]
\includegraphics[scale=0.25]{"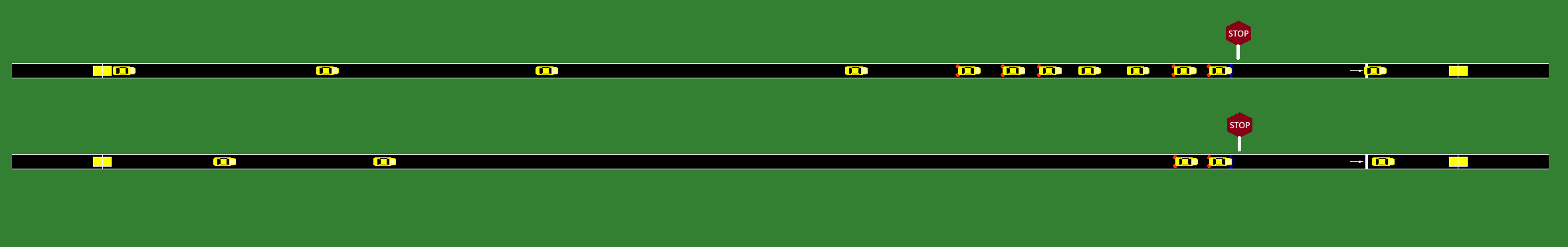"}
\caption{Screenshot of SUMO-based simulation experiment. The two equivalent routes with stop signs exhibit BPR-like dependence of mean travel time on flow. }
\label{Fig_SUMOscreen}
\end{figure}
\begin{figure}
\includegraphics[scale=0.6]{"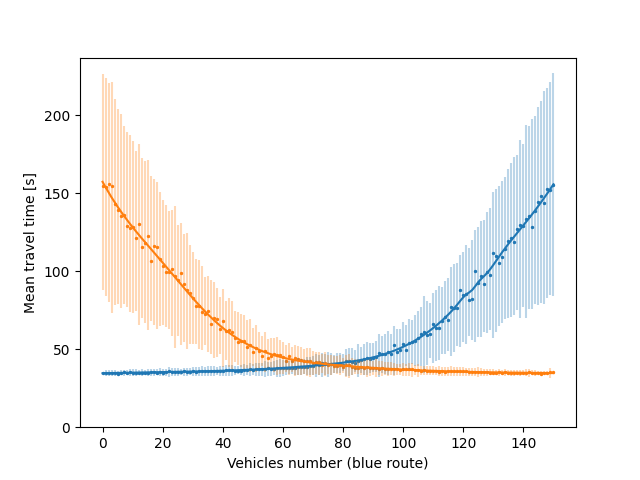"}
\caption{Dependence of average travel time on routes (blue - upper route, orange - lower route) on vehicle number in the SUMO-based system. The number of vehicles on orange route equals $150$ minus the number of vehicles on blue route. Error bars correspond to standard deviation. Data from $10$ indepent simulations.} 
\label{Fig_TRS}
\end{figure}

While the experiments in the main text clearly demonstrated the advanteges of randomized algorithms, they still used simplified flow-delay relations based on BPR functions. This experiment demonstrates the first step towards more realistic scenarios which is conducted as a microsimulation in SUMO, \cite{SUMO}. To do this, we designed a system of two routes with properties similar to the one used in the main experiments. We assumed that:

\begin{itemize}
\item There are two equivalent routes with stop signs, see Fig.  \ref{Fig_SUMOscreen}.
\item There are 150 vehicles, departing within 600 seconds.
\item Each vehicle (be it HDV or CAV) chooses/is assigned route, however it cannot choose the exact departure time. 
\item Vehicles arrive at start of routes with uniform probability within 10 min.
\item The travel time depends on congestion on each route, and is stochastic: a driver may get stuck at STOP for long if unlucky or may pass quickly. 
\item A day of travel is an independent SUMO run yielding actual private travel times for every driver and public means on routes. \item HDV utilities are given by \eqref{eq_utHDV} with $\overline{t_r} = \sum_{\iota = j-j_{avg}}^{j-1} t^{SUMO}_r (q_r(\iota))$, where $q_r(\iota)$ is the number of vehicles that chose route $r$ on day $\iota$ and $t_r^{SUMO}$ is the mean dependence of travel time on flow from Fig. \ref{Fig_TRS}. 
\item CAV utilities are based on \eqref{eq_utCAV} with CAV fleet credibilities based on \eqref{eq_cred} with $t_{r_i^{j}}^{j}$ replaced by $t_i^{j, SUMO}$, where $t_i^{j, SUMO}$ is the private travel time of driver $i$ in the SUMO simulation of day $j$. 
\item CAVs use the (known) dependece in Fig. \ref{Fig_TRS} for their algorithms.     
\end{itemize}

\begin{figure}
\includegraphics[scale=0.7]{"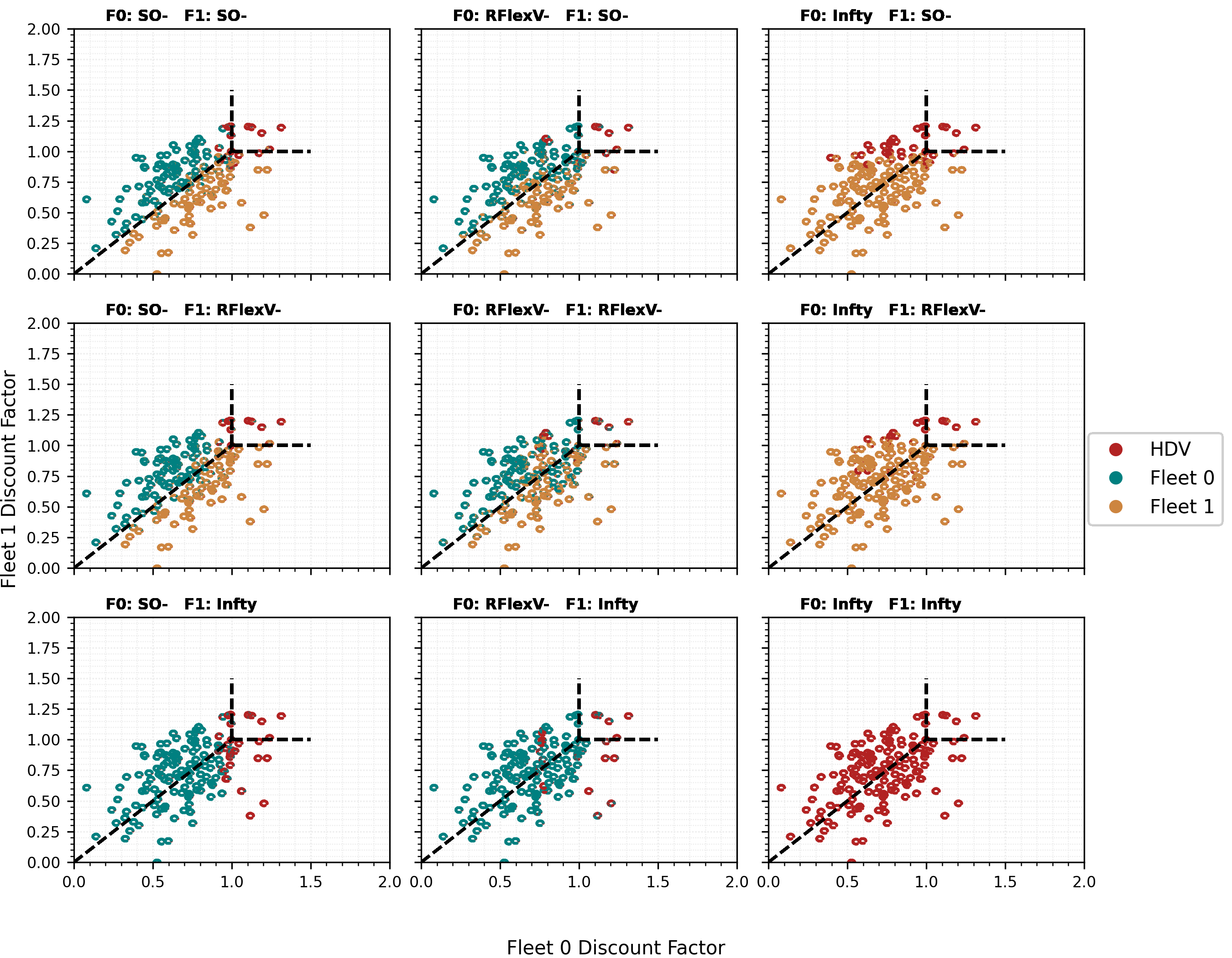"}
\caption{Modal split on days $1-200$ in the SUMO-based experiment. Each circle sector depicts the chosen modes of a driver (angle $0$ - day $1$, angle $3\pi/2 $ - day $200$). The advantage of randomised algorithms over SO- is clearly present in this setting as well.}
\label{Fig_SumoV}
\end{figure}

The experimental results shown in Fig. \ref{Fig_SumoV} demonstrate that:
\begin{itemize}
\item The randomized algorithms RFlexV- manages to achieve higher market share than baseline SO-. 
\item The results presented in this paper paper are likely to generalize to real-world scenarios. 
\end{itemize}

\end{document}